\documentclass[letterpaper,10pt]{article} 

\usepackage{osameet3} 

\usepackage{amsmath,amssymb,float}
\usepackage[colorlinks=true,bookmarks=false,citecolor=blue,urlcolor=blue]{hyperref} 

\begin{document}

\title{Stability analysis of generalized Lugiato-Lefever equation with lumped filter for Kerr soliton generation in anomalous dispersion regime}

\author{Nuo Chen,$^{*}$ Boqing Zhang,$^{*}$ Haofan Yang, Xinda Lu, Shiqi He, Yuhang Hu, Yuntian Chen, Xinliang Zhang and Jing Xu$^{\dagger}$}
\affiliation{School of Optical and Electronic Information $\&$ Wuhan National Laboratory for Optoelectronics,\\Huazhong University 
of Science and Technology, Wuhan 430074, China}
\email{$^{*}$These authors contributed equally to this work}
\email{$^{\dagger}$Corresponding author: jing\_xu@hust.edu.cn}
\copyrightyear{2021}

\begin{abstract}
We raise a detuning-dependent loss mechanism to describe the Kerr soliton formation dynamics when the lumped filtering operation is manipulated in anomalous group velocity dispersion regime, using stability analysis of Lugiato-Lefever equation.
\end{abstract}

\section{Introduction}
Optical frequency combs (OFCs), which consist of equidistant spectral lines, have revolutionized areas including precision spectroscopy, advanced frequency metrology, fiber-optic communications \cite{111}. Particularly, microresonator-based OFCs are promising to achieve highly compact, chip-scale integrated broadband comb sources at low power and microwave to Terahertz repetition rates. These advances can be attributed to the observation of temporal dissipative Kerr solitons (DKS) in microresonators, whose dynamics can be captured by the numerically solving a rigorous mapped Lugiato-Lefever equation (LLE) in nonlinear optics \cite{222}. Different dynamical behaviors of the system including Turing rolls, modulation instability (MI), breather soliton and single soliton, have been demonstrated in the experiments and agree well with stability analysis of LLE \cite{333}. On the other hand, the role of frequency dependent dissipation, i.e., spectral filtering, in parametric processes \cite{444} and further in the field of OFC attracted growing interests since MI can be induced by such an operation in normal group-velocity dispersion (GVD) regime. However, the impact of filtering in OFC in anomalous GVD regime has not been explored yet. In this work, we report stability analysis of LLE with lumped filtering on the formation of single soliton in anomalous GVD regime. According to the numerical simulation of LLE, it is found that the existence boundary of single soliton changes significantly from conventional LLE case. By introducing a detuning-dependent round trip loss, stability analysis of LLE with lumped filtering predicts identical behavior as numerical result, indicating the complex interaction between loss and Kerr-induced frequency shift. In addition, deterministic single soliton can be formed in this case with fixed pump scanning speed, avoiding complex scanning trajectory.

\section{Numerical simulations of GLLE with lumped filter}
Mathematically, the impact of lumped filtering in each round-trip is modelled by the convolution of intra-cavity field and filter function in Fourier domain, i.e., a generalized Lugiato-Lefever equation (GLLE) \cite{555}
\begin{equation}
    t_R \frac{\partial E(t,\tau)}{\partial t} =[-\alpha-i \delta-i \beta L  \frac{\partial^2}{\partial\tau^2}+i\gamma L|E|^2]E-f(\tau)\ast E+\sqrt{\kappa}E_{in}
\end{equation}\label{e1}
where $E(t,\tau)$ is the considered intracavity field envelope, $t$ the slow time and $\tau$ the fast time, $\delta$ is the frequency detuning between the external pump field $E_{in}$ and resonant mode. $\alpha$, $\beta$, $\gamma$ are the cavity intrinsic loss, the second-order dispersion coefficient and the Kerr nonlinear coefficient, respectively. Higher order dispersion effects are neglected. $\kappa$ is the coupling strength between the waveguide and the ring resonator, $L$ is perimeter of the ring, and $t_R$ is the round-trip time. $f(\tau)\ast E$ describes the convolution of field and filtering effect in time domain, in which $f(\tau)$ and $\widetilde{f}(\omega)$ are a Fourier transform pair. In general, $\widetilde{f}(\omega)$ can be taken as a Gaussian filter.

By employing a standard split-step Fourier method, we give soliton evolution processes on the $(\delta-E_{in}^2)$-map, as shown in Fig. 1. The pump power ranges from 0.1W to 2W at equal intervals sampling step and evolution at each pump is the superposition of 20 intracavity power traces. The power traces clearly reflect the soliton formation processes, and the region of distinct dynamical behaviours are revealed by comparing the relative intracavity power.
\begin{figure}[htbp]
    \centering
    \includegraphics[scale=0.4]{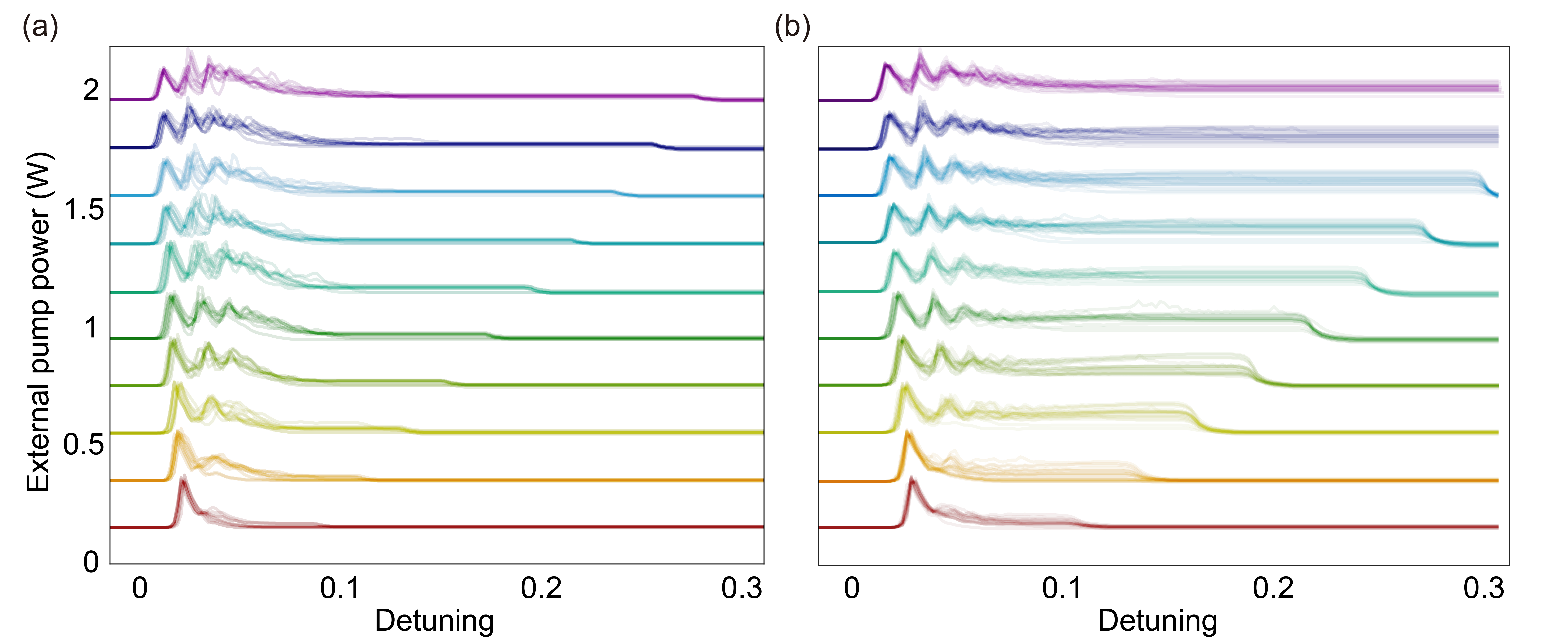}
    \caption{Superposition of intracavity power traces captured by numerical simulation of LLE in anomalous GVD regime, on a detuning-pump power map. In (a) the relative mode 10 is depleted in Fourier domain and in (b) is not. Parameters used are: $t_R=4.4ps$, $L=628\mu$m, $\alpha=2.4\times10^{-3}$, $\beta=-81ps^2/km$, $\gamma=1(W\cdot m)^{-1}$,$ \kappa=3.75\times10^{-4}$, and external pump $\left|E_{in}\right|^2$ is from 0.1W to 2W at equal intervals with 20 traces per sampling point of pump power.}
    \label{fig0}
\end{figure}
\section{Stability analysis of GLLE with lumped filter}
To carry out stability analysis of GLLE with lumped filter, the convolution term in Eq. (1) is replaced by a detuning-dependent loss term
\begin{equation}
    \frac{\partial E(t,\tau)}{\partial t} =[-\alpha(\delta)-i\delta-i\ \frac{\partial^2}{\partial\tau^2}+i|E|^2]E+E_{in}
\end{equation} \label{e2}
Note that Eq. (2) is normalized in order to draw conclusions about the universal nature of this kind of differential equations. Here $\alpha(\delta)$ gives the total energy decay, the power decay caused by filtering is treated as a loss related to detuning, in a linear form $\alpha\left(\delta\right)=\alpha_0+C\delta$, where $C$ indicates the power attenuation induced by lumped filtering. By exploiting stability analysis \cite{333}, it is necessary to find all the equilibria by setting all the differential terms with respect to time to zero, yielding
\begin{equation}
    [-\alpha(\delta)-i\delta+i|E|^2]E+E_{in}=0
\end{equation} \label{e3}
After mathematical deduction, in $(\delta-E_{in}^2)$-plane, the lines of two equilibria boundaries $E_{in-}^2(\delta)$ and $E_{in+}^2(\delta)$ can be expressed in terms of detuning $\delta$
\begin{equation}
    E_{in-}^2(\delta)=\frac{2\delta+\sqrt{\delta^2-3\alpha^2(\delta)}}{3}\left[\alpha^2(\delta)+\left(\frac{\sqrt{\delta^2-3\alpha^2(\delta)}-\delta}{3}\right)^2\right] \tag{4a}
\end{equation} \label{e4a}
\begin{equation}
    E_{in+}^2(\delta)=\frac{2\delta-\sqrt{\delta^2-3\alpha^2(\delta)}}{3}\left[\alpha^2(\delta)+\left(\frac{\sqrt{\delta^2-3\alpha^2(\delta)}+\delta}{3}\right)^2\right] \tag{4b}
\end{equation} \label{e4b}
The boundaries of bifurcations Eq. (4a) and Eq. (4b) are sketched in Fig. 2, represented by white dashed lines. Figure 2 illustrates the consistency of simulation results and the boundaries determined by stability feature of GLLE, it supports the fact that the detuning-dependent loss mechanism is valid.

Interestingly, numerical simulation shows that the single soliton generation probability is significantly increased even close to 1 when the lumped filtering operation is taking effect, so we statistically transfer the power traces in Fig. 1 into a heat map so that it reports the bright soliton existence regime as well as the probability of single soliton generation (see color bar in Fig. 2). The pump power sampling is relatively smoother in Fig. 2, ranges from 0W to 2W per 0.01W, with 100 simulations per pump power.
\begin{figure}[htbp]
    \centering
    \includegraphics[scale=0.53]{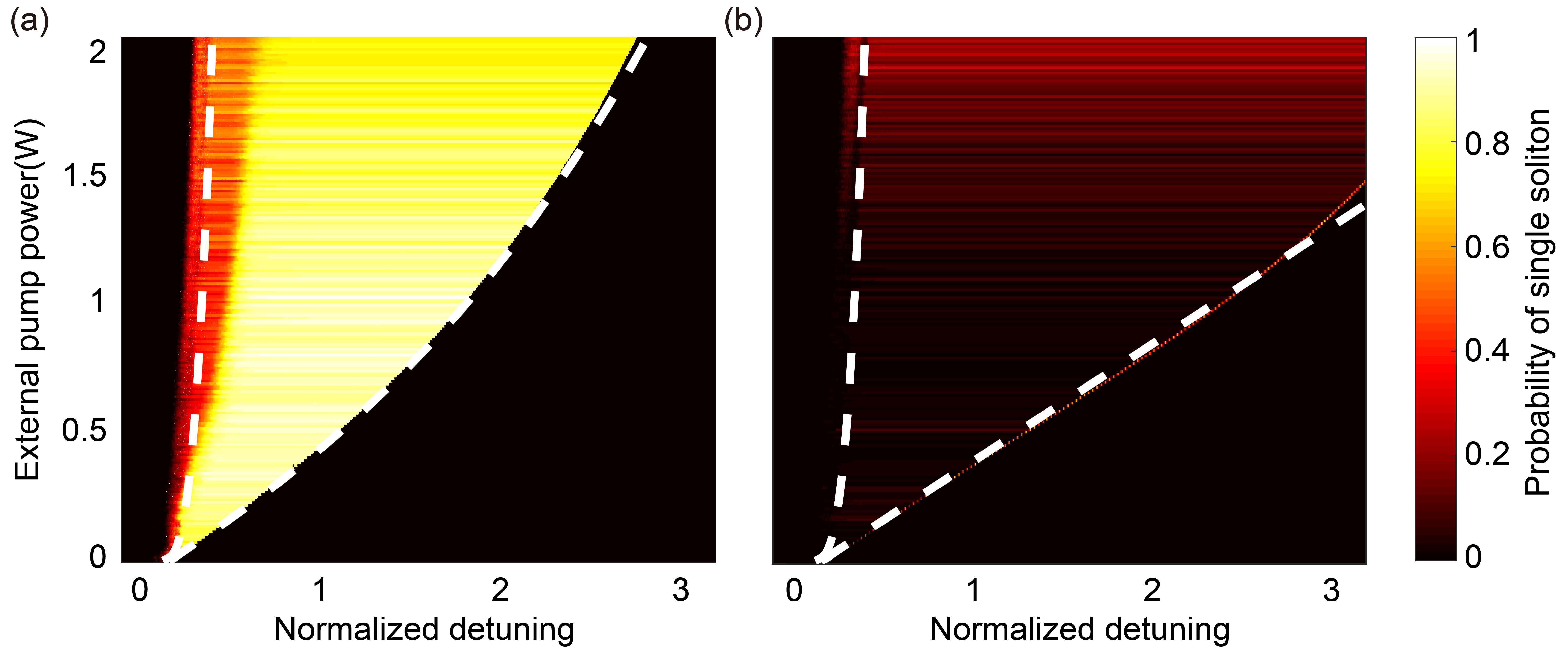}
    \caption{Probability of single soliton formation on the pump power-detuning map with (a) and without (b) filtering, as well as the comparison of the stability boundaries on a normalized detuning and pump power map. White dashed lines represent the higher/lower boundary of the bright soliton region determined by Eq. (4a) and Eq. (4b). Parameters used are same as Fig. 1, besides the sampling points of pump power is ranging from 0 to 2W per 0.01W, and we run 100 simulations per pump power sampling point. In (a) lumped filtering induced power attenuation coefficient is $C=2.8\times10^{-3}$.}
    \label{fig1}
\end{figure}
\section{Discussion}
We give a possible physical picture to expound the studied filtering effect on boundaries of the soliton region. In anomalous GVD regime, consider the relationship between a pump light and a resonance mode of the microring resonator, $\delta$ indicates the detuning between them, as the previous mathematical analysis. A high-quality factor (Q-factor) resonator can accumulate enormous energy, therefore, due to the Kerr nonlinear effect (neglecting the other effects), the resonance peak will offset from its origin center frequency, i.e. red-detuning or blue-detuning. In a dynamic process of generating Kerr optical frequency combs, the pump light scan crosses the resonance peak of the micro-ring resonator. At the beginning, pump frequency nears the resonance peak, and moves faster than it in the frequency domain, energy begins to accumulate in the ring, Kerr combs arises, Turing rolls or MI emerge. At this period the filter may does not work (in case of Turing rolls, at aimed frequency it has no power) or randomly dissipates energy (in case of MI). Then, at the moment when the pump field is aligned with the offset center of resonance peak, the intracavity energy reaches the maximum. As the pump light continues to detune, it exceeds the center frequency of the resonance peak, makes it detune in the other direction. From now on, the system gets to the soliton region and filtering effect works well. At this stage, the intracavity power begins to dissipate spontaneously, bright solitons generate (including the single-soliton/multi-solitons/breathers), the filtering mechanism accelerates the process of dissipation, makes the resonance peak return to the origin center frequency faster. As it returns faster, the filtering effect becomes stronger, a positive feedback forms, this feedback effect explains the lumped filtering induced loss term is a function of detuning. In the end, solitons maintenance time becomes shorter since energy dissipates faster, in $(\delta-E_{in}^2)$-plane, this change is manifested as the lower boundary of solitons $E_{in-}^2(\delta)$ uplifts.

In summary, we have theoretically investigated the stability of GLLE and proposed a detuning-dependent loss mechanism to describe the formation of solitons. Dynamics are studied using numerical simulation of LLE, and we are capable to predict the bright solitons generation regime by exploiting the stability analysis of GLLE in anomalous GVD regime.

This work is supported by the National Key Research and Development Program of China (2019YFB2203103), National Natural Science Foundation of China (NSFC) (no. 61775063, no. 11874026, and no. 61735006).


\begin{thebibliography}{99} 

\bibitem{111} T. J. Kippenberg, A. L. Gaeta, M. Lipson and M. L. Gorodetsky, ''Dissipative Kerr solitons in optical microresonators,'' Science, 361(6042): eaan8083 (2018).

\bibitem{222} S. Coen, H. G. Randle, T. Sylvestre and M. Erkintalo, ''Modeling of octave-spanning Kerr frequency combs using a generalized mean-field Lugiato–Lefever model,'' Optical Letters, 38(1): 37-39 (2013).

\bibitem{333} C. Godey, I. V. Balakireva, A. Coillet and Y. K. Chembo, ''Stability analysis of the spatiotemporal Lugiato-Lefever model for Kerr optical frequency combs in the anomalous and normal dispersion regimes,''  Physical Review A, 89: 063814 (2014).

\bibitem{444} H. Hu, L. Zhang, C. Zhang, Y. Chen, J. Xu and X. Zhang, ''Lumped dissipation induced quasi-phase matching for broad and flat optical parametric processes,'' IEEE Photonics Journal, 11(6): 1-8 (2019).

\bibitem{555} A. M. Perego, S. K. Turitsyn and K, ''Gain through losses in nonlinear optics,'' Light: Science \& Applications, 7: 43 (2018).

\end{thebibliography}
\end{document}